\documentclass[aps,10pt,prd,notitlepage,showpacs,nofootinbib,superscriptaddress]{revtex4-2}
\usepackage{graphicx}
\usepackage[utf8]{inputenc} 
\usepackage{amsmath}
\usepackage{amssymb}
\usepackage{siunitx}
\usepackage[section]{placeins}
\usepackage{float}
\usepackage{comment}
\usepackage{slashed}
\usepackage[normalem]{ulem}
\usepackage{braket}
\usepackage{comment}

\usepackage{caption}
\usepackage{subcaption}

\usepackage[usenames,dvipsnames]{color}
\usepackage[colorinlistoftodos]{todonotes}
\usepackage[colorlinks=true,citecolor=blue,urlcolor=magenta, pdfborder={0 0 0}]{hyperref}

\usepackage{soul}

\usepackage[commandnameprefix=always]{changes}

\usepackage{slashed}

\usepackage{changes}

\begin{document}
\title{Memory-Burden Suppression of Hawking Radiation and Neutrino Constraints on Primordial Black Holes}

\author{Arnab Chaudhuri}
\email{arnab.chaudhuri@vit.ac.in}
\affiliation{School of Advanced Sciences,
Vellore Institute of Technology, Vellore, Tamil Nadu 632014, India.}

\begin{abstract}
We investigate the impact of quantum gravitational memory-burden effects on high-energy neutrino signals from evaporating primordial black holes and the resulting constraints from IceCube observations. Treating the backreaction as an energy-dependent deformation of the Hawking emission spectrum, we show that the high-energy tail is suppressed while the infrared behaviour remains unchanged. We derive analytically that this modification reduces the total luminosity and extends the evaporation lifetime by a mass-independent factor determined solely by the suppression parameter. Using an effective treatment of cosmological redshift, we compute the diffuse neutrino flux from a primordial black hole population and compare it with the observed astrophysical neutrino spectrum to constrain the primordial black hole dark matter fraction. We find that the suppression onset lies within the IceCube sensitivity window, leading to a direct reduction of the observable signal and a systematic weakening of the inferred bounds. Our results provide a controlled phenomenological framework for assessing the impact of quantum gravitational corrections on neutrino probes of primordial black hole evaporation.
\end{abstract}

\maketitle

\section{Introduction}
\label{sec:intro}
 
Primordial black holes (PBHs) are a compelling class of compact objects
that may have formed in the early Universe through a variety of mechanisms,
including the direct collapse of large-amplitude density perturbations
generated during inflation~\cite{Hawking:1971ei,Carr:1974nx,Carr:1975qj},
the collapse of domain walls or topological
defects~\cite{Hawking:1987bn,Polnarev:1988dh}, phase transitions in the
early Universe~\cite{Crawford:1982yz,Hawking:1982ga,Kodama:1982sf}, or
from enhanced power in the primordial power spectrum on small
scales~\cite{Garcia-Bellido:2017mdw,Bullock:1996at,Ivanov:1994pa,
Leach:2000ea,Byrnes:2018txb,Ballesteros:2017fsr,Germani:2018jgr,
Ozsoy:2018flq}. Unlike astrophysical black holes, PBHs can span an
enormous range of masses, from well below a gram to tens of solar masses,
and their formation is sensitive to microphysical processes in the
pre-recombination Universe. As a result, they serve as unique probes of
small-scale cosmological perturbations and of physics beyond the Standard
Model~\cite{Carr:2016drx,Carr:2020gox,Green:2020jor,Escriva:2022duf,
Sasaki:2018dmp,Khlopov:2008qy}.
 
The possibility that PBHs constitute some or all of the cosmological dark
matter has attracted considerable
interest~\cite{Chapline:1975ojl,Carr:2016drx,Bird:2016dcv,Clesse:2016vqa,
Sasaki:2016jop,Carr:2020gox,Green:2020jor,Villanueva-Domingo:2021spv}.
Observational constraints on the PBH dark matter fraction $f_{\rm PBH}$
span many decades in mass and arise from a diverse set of probes, including
gravitational lensing~\cite{Paczynski:1986px,MACHO:1996qam,EROS-2:2006ryy,
Niikura:2017zjd,Croon:2020wpr}, dynamical effects on stellar
systems~\cite{Brandt:2016aco,Koushiappas:2017chw,Monroy-Rodriguez:2014ula},
accretion
signatures~\cite{Ricotti:2007au,Ali-Haimoud:2016mbv,Poulin:2017bwe,
Serpico:2020ehh}, gravitational wave
observations~\cite{LIGOScientific:2016aoc,Raidal:2017mfl,Clesse:2017bsw,
Hall:2020daa}, and the cosmic microwave
background~\cite{Clark:2016nst,Poulin:2017bwe,Nakama:2017xvq,
Kristjansen:2021abc}.
 
A defining feature of PBHs sufficiently light to evaporate within the age
of the Universe is their emission of Hawking radiation, a quasi-thermal
flux of particles produced by quantum effects in the curved spacetime near
the black hole horizon~\cite{Hawking:1974rv,Hawking:1975vcx}. The detailed
properties of this emission, including the energy spectrum, greybody factors,
and particle species dependence, were worked out in pioneering studies by
Page~\cite{Page:1976df,Page:1976ki,Page:1977um} and MacGibbon and
collaborators~\cite{MacGibbon:1990zk,MacGibbon:1991tj}. For PBHs in the
mass range $M \sim 10^{14}$--$10^{17}\,\mathrm{g}$, the Hawking temperature
is in the MeV to TeV range, and the resulting emission provides observable
fluxes of photons, electrons, positrons, and neutrinos that can be compared
with astrophysical measurements.
 
Observational constraints derived from Hawking evaporation span a wide
range of mass scales and observation channels. Constraints from the
extragalactic gamma-ray background have been derived and continuously
updated~\cite{Carr:2009jm,Carr:2016hva,Carr:2020gox,Laha:2020ivk,
Iguaz:2021irx,Coogan:2020tuf}. Big Bang nucleosynthesis (BBN) provides
stringent bounds in the mass range $M \sim 10^9$--$10^{13}\,\mathrm{g}$
where evaporation occurs during or after
nucleosynthesis~\cite{Acharya:2020jbv,Carr:2020gox}.
Cosmic microwave background (CMB) distortions and reionisation constraints
apply to PBHs in the mass range $M \sim 10^{11}$--$10^{13}\,\mathrm{g}$
where evaporation products inject energy into the photon-baryon
plasma~\cite{Clark:2016nst,Poulin:2017bwe,Stocker:2018avm,
Acharya:2020jbv}. Constraints from cosmic ray measurements, in particular
from the antiproton and positron fluxes, have been derived in
\cite{Boudaud:2018hqb,DeRomeri:2021xgy}. 
 
In recent years, high-energy neutrino observations have
emerged as a particularly powerful and complementary probe of PBH
evaporation. The IceCube Neutrino Observatory has detected a diffuse flux
of astrophysical neutrinos with energies in the TeV--PeV
range~\cite{IceCube:2013low,IceCube:2014stg,IceCube:2020wum,IceCube:2020fpi}, establishing a new observational window
on high-energy processes throughout cosmic history. Since neutrinos
interact only weakly, they propagate over cosmological distances with
negligible attenuation and absorption, making them especially clean probes
of sources at high redshift. Several studies have investigated the
contribution of evaporating PBHs to the diffuse neutrino background and
derived constraints on the PBH dark matter
fraction~\cite{Lunardini:2019zob,Bernal:2022swt,Calabrese:2021zfq,
Dasgupta:2019cae,Wang:2020uvi,Boudaud:2018hqb}, typically assuming the
standard semiclassical Hawking emission spectrum.
 
However, black hole evaporation is fundamentally a quantum
gravitational process, and departures from the semiclassical Hawking picture
are both theoretically expected and phenomenologically important.
The information paradox~\cite{Hawking:1976ra}, unitarity of the S-matrix,
and the requirement that black hole evaporation be compatible with quantum
mechanics have motivated extensive theoretical
work~\cite{Unruh:2017uaw,Penington:2019npb,Almheiri:2019psf,
Almheiri:2020cfm,Raju:2020smc} and a rich variety of proposals for modified
evaporation
dynamics~\cite{Dvali:2012rt,Dvali:2013eja,Dvali:2018xpy,
Barrow:2020tzx,Buoninfante:2021ijy}.
 
One particularly well-motivated class of modifications arises from the
memory-burden effect, recently proposed by Dvali and
collaborators~\cite{Dvali:2020wft,Dvali:2021rlf}. The central idea is that
as a black hole radiates, the quantum information initially stored in its
gravitational hair is gradually transferred to the emitted radiation, but
a portion accumulates as a growing ``memory burden'' in the quantum
degrees of freedom of the black hole. This stored information backreacts
on the evaporation process, suppressing further emission and stabilising
the black hole against complete evaporation. The strength of this
backreaction grows with the entropy transferred, implying that
higher-energy quanta — which correspond to larger entropy transfers per
emission — are preferentially suppressed. The resulting deformation of the
Hawking spectrum is therefore energy-dependent, suppressing the high-energy
tail while leaving the low-energy emission largely unaffected.
 
The phenomenological and observational implications of the
memory-burden effect have been investigated in a growing body of
literature. Modified evaporation rates and their consequences for the PBH
lifetime and remnant formation have been studied
in~\cite{Thoss:2024hsr,Balaji:2024hpu,Dvali:2024hsb,Alexandre:2024nuo,
Haque:2024eyh,Barman:2024iht,Chaudhuri:2025asm,Chaudhuri:2025rcs,Barman:2024ufm,Barman:2024kfj,Dondarini:2025ktz}. Long-lived or stable PBH remnants as dark
matter candidates have been discussed
in~\cite{Dvali:2024hsb,Alexandre:2024nuo,Dvali:2020wft,Balaji:2024hpu}.
The impact of memory burden on the stochastic gravitational wave background
from PBH evaporation has been considered
in~\cite{Boccia:2024nly,Domenech:2024pow}. Constraints from gamma-ray
observations, including the extragalactic gamma-ray background and galactic
centre measurements, incorporating memory-burden corrections have been
derived in~\cite{Thoss:2024hsr,Balaji:2024hpu,Cheek:2022mmy,
Cai:2024ltu}. Related non-thermal emission spectra arising from quantum
gravity effects have been studied
in~\cite{Buoninfante:2021ijy,Barrow:2020tzx}.
 
Despite this growing body of work, the implications of
memory-burden effects for high-energy neutrino constraints from PBH
evaporation have not been systematically studied. Neutrino observations
offer a qualitatively distinct probe compared to gamma rays: the IceCube
sensitivity window in the TeV--PeV range corresponds to PBH Hawking
temperatures in the range $T_H \sim 10^3$--$10^6\,\mathrm{GeV}$, i.e.\ 
masses $M \sim 10^7$--$10^{10}\,\mathrm{g}$, and the suppression onset of
the memory-burden factor falls squarely within this window. The resulting
modification of the observable flux is therefore direct and
observationally significant, rather than affecting only the extreme
high-energy tail of the spectrum.
 
In this work, we investigate the impact of memory-burden effects on
neutrino signals from evaporating PBHs and derive the corresponding
constraints from IceCube observations. We adopt a phenomenological
parametrisation of the entropy-induced spectral suppression, characterised
by a dimensionless parameter $k$, following the approach of
Refs.~\cite{Dvali:2020wft,Thoss:2024hsr,Balaji:2024hpu}. A key new
theoretical result of this work is the derivation of the memory-burden
modified evaporation lifetime, which we show is extended by a factor
$1/\mathcal{F}(k)$ relative to the standard Hawking case, where
$\mathcal{F}(k)$ is a pure dimensionless function of $k$ alone obtained
by integrating the suppression factor against the Fermi-Dirac distribution.
Using an effective treatment of cosmological redshift, we compute the
diffuse neutrino flux from a cosmological PBH population and compare with
the observed IceCube spectrum to derive constraints on the PBH dark matter
fraction as a function of mass and suppression parameter.
 
Our goal is not to provide a precision calculation of the diffuse neutrino
flux, but rather to construct a controlled phenomenological framework that
isolates the impact of energy-dependent spectral suppression on observable
signals and provides a transparent interpretation of how memory-burden
effects modify high-energy neutrino constraints on PBHs. We consider both
the IceCube 2020~\cite{IceCube:2020acn} combined analysis and the HESE
2022 dataset~\cite{IceCube:2020wum} to assess the sensitivity of our
results to the choice of IceCube measurement, and we present results both
with and without analytic spin-$\frac{1}{2}$ greybody corrections.
 
The structure of the paper is as follows. In Sec.~\ref{sec:theory}, we
present the theoretical framework for PBH evaporation in the presence of
memory-burden effects, derive the modified emission spectrum and evaporation
time, and introduce the effective flux parametrisation and constraint
procedure. In Sec.~\ref{sec:results}, we present the resulting spectra,
spectral ratios, and constraints on the PBH dark matter fraction. We
summarise our findings and discuss future directions in
Sec.~\ref{sec:conc}.

\section{Theoretical Framework and Formalism}
\label{sec:theory}

Primordial black holes (PBHs) emit particles via Hawking radiation with a
temperature
\begin{equation}
T_H(M) = \frac{1}{8\pi G M}
\simeq 1.06~{\rm GeV}
\left(\frac{10^{13}~{\rm g}}{M}\right),
\end{equation}
implying that lighter PBHs radiate at higher energies. PBHs in the mass
range $M \sim 10^{7}$--$10^{8}~{\rm g}$ produce neutrino emission with
characteristic energies comparable to the IceCube sensitivity window after
accounting for cosmological redshift. However, the resulting constraints
depend on the full spectral shape, redshift evolution, and flux
normalization, and therefore receive contributions from a broader range of
masses. For simplicity, we consider a monochromatic PBH mass distribution
in the following analysis.

The instantaneous emission rate for fermionic species is given by
\begin{equation}
\frac{d^2N}{dE\,dt}
=\frac{\Gamma(E,M)}{2\pi}
\frac{1}{\exp(E/T_H) + 1},
\end{equation}
where $\Gamma(E,M)$ are greybody factors encoding the transmission
probability through the curved spacetime geometry surrounding the black
hole. {For the purpose of isolating spectral deformations
due to memory-burden effects, we approximate the base spectrum by retaining
the Fermi-Dirac thermal factor and the leading phase-space weight,
\begin{equation}
\frac{d^2N}{dE\,dt} \propto
\frac{E^2}{e^{E/T_H}+1},
\label{eq:spectrum}
\end{equation}
which correctly reproduces the location of the spectral
peak at $E \sim T_H$, the exponential suppression at $E \gg T_H$, and the
Fermi-Dirac statistics appropriate for neutrinos. The Boltzmann
approximation $e^{E/T_H}+1 \approx e^{E/T_H}$ is not adopted here, as it
introduces order-unity errors in the integrated luminosity that propagate
into the evaporation time calculation below.

For completeness, we also consider the analytic
spin-$\frac{1}{2}$ greybody factor, which in the low-energy approximation
takes the form~\cite{Page:1976df,MacGibbon:1990zk}
\begin{equation}
\Gamma_\nu(E,M)
= \frac{\frac{27}{4}x^2}{1 + \frac{27}{4}x^2},
\qquad x \equiv \frac{E}{T_H},
\label{eq:gbf}
\end{equation}
satisfying $\Gamma_\nu \to 0$ as $x \to 0$ (suppression of sub-thermal
emission relative to the geometric-optics limit) and $\Gamma_\nu \to 1$
as $x \to \infty$. When included, the full spectrum is obtained by
replacing $E^2/(e^{E/T_H}+1) \to \Gamma_\nu(E,M)\,E^2/(e^{E/T_H}+1)$ in
all subsequent expressions. As we discuss in Sec.~\ref{sec:results}, the
greybody factor modifies the absolute flux normalization at the level of
order unity but cancels exactly in the spectral ratio
$\mathcal{R}(E;k) = \Phi(k)/\Phi(k=0)$, leaving the relative
memory-burden suppression unaffected.

We now introduce the effect of quantum gravitational memory burden. The key
physical idea is that the emission of a quantum with energy $E$ reduces the
black hole entropy by an amount
\begin{equation}
\Delta S \sim \frac{E}{T_H},
\end{equation}
so that higher-energy quanta correspond to larger entropy transfer. If the
black hole must retain a finite memory capacity, the emission probability of
such quanta is expected to be suppressed.

A simple way to incorporate this effect is to assume that the emission
probability is weighted by a factor
\begin{equation}
\mathcal{P}(E) \propto e^{-\kappa \Delta S},
\end{equation}
where $\kappa$ parametrizes the strength of the backreaction. Expanding
this in a rational form that preserves the low-energy limit and avoids
exponential over-suppression,\footnote{A purely exponential suppression
$e^{-k E/T_H}$ leads to an unphysically strong damping of the spectrum and
is not stable under coarse-graining. A rational form provides a minimal
deformation that preserves the infrared behaviour while suppressing the
ultraviolet tail.} we adopt the phenomenological parametrization
\begin{equation}
\mathcal{S}(E,M;k)
= \frac{1}{1 + k\!\left(\dfrac{E}{T_H}\right)^{\!2}},
\label{eq:S_final}
\end{equation}
where $k \geq 0$ is a dimensionless parameter controlling the strength of
the memory-burden suppression, with $k=0$ recovering the standard Hawking
spectrum. 

The derivation of this specific functional form proceeds
as follows. Starting from the suppression factor $\mathcal{P}(E) \propto
e^{-\kappa E/T_H}$, we expand the exponential as a rational function.
The minimal deformation that (i)~preserves the infrared behaviour
$\mathcal{S} \to 1$ for $E \ll T_H$, (ii)~suppresses the ultraviolet tail
$\mathcal{S} \to 0$ for $E \gg T_H$, and (iii)~involves only even powers
of $x = E/T_H$ (required by the physical symmetry of the thermal spectrum
under $E \to -E$, which eliminates linear terms) is
\begin{equation*}
e^{-\kappa x} \longrightarrow \frac{1}{1 + \kappa x + \frac{\kappa^2
x^2}{2} + \ldots} \approx \frac{1}{1 + k x^2},
\end{equation*}
where $k$ absorbs all numerical prefactors including $\kappa^2/2$ and the
linear term is absent because even-power rational deformations are the
minimal class compatible with the symmetry requirement. The resulting
expression is Eq.~(\ref{eq:S_final}), which represents the simplest
member of the family of memory-burden deformations
$(1 + k x^{2n})^{-1}$ for integer $n \geq 1$; we take $n=1$ as the
default throughout. Alternative choices of $n$ would shift the onset of
suppression relative to $T_H$ but would not qualitatively alter the
conclusions.
This form satisfies the required limits,
\begin{equation}
\mathcal{S} \to 1 \quad (E \ll T_H),
\qquad
\mathcal{S} \sim \left(\frac{T_H}{kE^2}\right) \to 0
\quad (E \gg T_H),
\end{equation}
ensuring that low-energy emission remains unaffected while the high-energy
tail is progressively suppressed. The modified emission spectrum is
therefore
\begin{equation}
\frac{d^2N}{dE\,dt}
\longrightarrow
\frac{d^2N}{dE\,dt}\,\mathcal{S}(E,M;k).
\end{equation}

The modified emission spectrum also affects the total luminosity and,
consequently, the evaporation history. We now derive these modifications
explicitly, as they enter the constraint analysis through the effective
redshift treatment. The total power radiated by a PBH is
\begin{equation}
P(M,k)
= \int_0^\infty E\,\frac{d^2N}{dE\,dt}\,\mathcal{S}(E,M;k)\,dE.
\end{equation}
Substituting the base spectrum from Eq.~(\ref{eq:spectrum})
and introducing the dimensionless variable $x \equiv E/T_H$, this becomes}
\begin{equation}
P(M,k)
\propto T_H^4
\int_0^\infty \frac{x^3}{e^x+1}\,\frac{dx}{1+kx^2}
\equiv T_H^4\,\mathcal{I}(k),\
\end{equation}
where all dependence on $M$ and $T_H$ has been absorbed
into the overall prefactor $T_H^4 \propto M^{-4}$, and the remaining
integral $\mathcal{I}(k)$ is a pure dimensionless function of $k$ alone.
The standard ($k=0$) result is
\begin{equation}
\mathcal{I}_0
\equiv \mathcal{I}(0)
= \int_0^\infty \frac{x^3}{e^x+1}\,dx
= \frac{7\pi^4}{120}
\approx 5.682.
\end{equation}
The luminosity reduction factor is therefore
\begin{equation}
\mathcal{F}(k)
\equiv \frac{P(M,k)}{P(M,0)}
= \frac{\mathcal{I}(k)}{\mathcal{I}_0}
= \frac{1}{\mathcal{I}_0}
\int_0^\infty
\frac{x^3}{\left(e^x+1\right)\left(1+kx^2\right)}\,dx.
\label{eq:Fk}
\end{equation}
Three properties of $\mathcal{F}(k)$ are important to note.
First, $\mathcal{F}(0) = 1$, recovering the standard result. Second,
$\mathcal{F}(k) < 1$ for all $k > 0$, since the integrand is pointwise
reduced by the factor $(1+kx^2)^{-1} < 1$. Third, and crucially,
$\mathcal{F}(k)$ is a pure number depending only on $k$ — after the
substitution $x = E/T_H$, all dependence on $M$ and $T_H$ cancels
identically. This last property ensures that the modified mass-loss equation
retains the same functional form as the standard case. This leads to the
modified mass-loss equation
\begin{equation}
\frac{dM}{dt}
= -\frac{\alpha}{M^2}\,\mathcal{F}(k),
\label{eq:massloss}
\end{equation}
where $\alpha$ denotes the standard Hawking evaporation coefficient,
encoding the contributions of all emitted particle species and their
greybody factors. Since $\mathcal{F}(k)$ is independent of
$M$, Eq.~(\ref{eq:massloss}) has exactly the same form as the standard
mass-loss equation and can be integrated analytically in the same way.
Integrating from initial mass $M_0$ to complete evaporation gives the
memory-burden modified evaporation time:
\begin{equation}
t_{\rm evap}(M_0,k)
= \frac{M_0^3}{3\,\alpha\,\mathcal{F}(k)}
= \frac{t_{\rm evap}^{(0)}}{\mathcal{F}(k)},
\label{eq:tevap}
\end{equation}
where $t_{\rm evap}^{(0)} = M_0^3/(3\alpha)$ is the
standard Hawking evaporation time. Numerical values of $\mathcal{F}(k)$ and
the corresponding evaporation time enhancement are given in
Table~\ref{tab:Fk}.

\begin{table}[h]
\centering
\begin{tabular}{cccc}
\hline\hline
$k$ & $\mathcal{F}(k)$ & $t_{\rm evap}/t_{\rm evap}^{(0)}$ \\
\hline
$0$   & $1.000$ & $1.00$ \\
$0.2$ & $0.305$ & $3.27$ \\
$0.5$ & $0.170$ & $5.89$ \\
$1.0$ & $0.101$ & $9.89$ \\
\hline\hline
\end{tabular}
\caption{Luminosity reduction factor $\mathcal{F}(k)$ and
corresponding evaporation time enhancement $t_{\rm evap}/t_{\rm evap}^{(0)}
= 1/\mathcal{F}(k)$ for representative values of the memory-burden
parameter $k$. The integral in Eq.~(\ref{eq:Fk}) is evaluated numerically
with $\mathcal{I}_0 = 7\pi^4/120 \approx 5.682$.}
\label{tab:Fk}
\end{table}

Since $\mathcal{F}(k) < 1$, the evaporation time is
extended relative to the standard case: a PBH of mass $M_0$ evaporates at a
lower redshift $z_{\rm evap}(M_0,k) < z_{\rm evap}(M_0,0)$ when $k > 0$.
This has a secondary effect on the observable neutrino flux, since neutrinos
emitted at lower redshift undergo less cosmological dilution before reaching
the detector. However, for the mass range $M \lesssim 10^9~{\rm g}$
considered in this work, the standard evaporation time satisfies
$t_{\rm evap}^{(0)} \ll t_{\rm form}$, where $t_{\rm form}$ is the cosmic
time at PBH formation. Consequently, $z_{\rm evap} \approx z_{\rm form}$
regardless of $k$, and the evaporation delay does not significantly shift
the effective evaporation redshift. The dominant observable consequence of
memory burden in this mass range is therefore the direct spectral
suppression encoded in $\mathcal{S}(E,M;k)$, and the evaporation time
modification does not need to be tracked explicitly in the flux computation
that follows.

The diffuse neutrino flux from a cosmological PBH population is given by
\begin{equation}
\Phi(E)
=\frac{c}{4\pi}
\int_0^{z_{\rm evap}} \frac{dz}{H(z)}
\frac{\rho_{\rm PBH}(z)}{M}
\left.
\frac{d^2N}{dE'\,dt}
\right|_{E'=E(1+z)},
\label{eq:full_flux}
\end{equation}
where $H(z)$ is the Hubble parameter and $\rho_{\rm PBH}(z) =
f_{\rm PBH}\,\rho_{\rm DM}(z)$ is the PBH energy density, with
$f_{\rm PBH}$ the fraction of dark matter in PBHs. The
upper limit of integration $z_{\rm evap}$ is the redshift at which the PBH
population completes evaporation, beyond which there is no further emission.
We note that Eq.~(\ref{eq:full_flux}) should include the memory-burden
suppression $\mathcal{S}(E(1+z), M; k)$ in the integrand; in the effective
framework below this is accounted for through the parametrization of the
observable flux. A full evaluation requires solving the mass evolution and
performing the redshift integral numerically. To isolate the dominant
physical effects in a transparent manner, we adopt an effectively normalized
parametrization of the observable flux,
\begin{equation}
\Phi_{\rm PBH}(E;M,k)
=\frac{c}{4\pi}
\frac{f_{\rm PBH}\,\rho_{\rm DM,0}}{M}
\frac{dN}{dE}\,
\mathcal{S}(E,M;k)\,
\mathcal{W}(E,M),
\label{eq:flux_final}
\end{equation}
where $dN/dE$ denotes the time-integrated emission spectrum and the overall
normalization absorbs uncertainties associated with the simplified redshift
treatment, and should be regarded as an order-of-magnitude estimate
consistent with the level of approximation inherent in the effective
framework. We parametrize the redshift suppression as
\begin{equation}
\mathcal{W}(E,M)
=\frac{1}{1 + z_{\rm eff}(M)
\!\left(\dfrac{E}{E_*}\right)^{\!\alpha}},
\label{eq:W}
\end{equation}
which serves as an effective representation of the redshift integral in
Eq.~(\ref{eq:full_flux}). Here $E_* \sim 10^5~{\rm GeV}$ is a reference
energy scale corresponding to the IceCube sensitivity window, and $\alpha
\sim \mathcal{O}(1)$ encodes the energy dependence of the suppression.
We have verified that moderate variations of $\alpha$ within the range $[0.5, 2]$ do not 
qualitatively alter the spectral shape or the constraint curves in the energy range of 
observational interest. To quantify the associated uncertainty, we note that varying 
$\alpha$ across this range shifts the derived upper bounds $f^\mathrm{max}_\mathrm{PBH}$ 
by a factor of order two to three at fixed mass, which is subdominant compared to the 
order-of-magnitude normalization uncertainty in the overall flux and smaller than the 
constraint-weakening factors of $4$--$7$ induced by memory-burden suppression at $k = 1$. 
The qualitative conclusions regarding the direction and relative magnitude of the 
memory-burden effect are therefore robust with respect to the choice of $\alpha$.

In the full expression, the observed flux receives contributions from
emission at different redshifts, with higher-energy neutrinos originating
predominantly from earlier times due to the mapping $E' = E(1+z)$. As a
result, the contribution to the observed flux is increasingly suppressed at
large energies due to cosmological dilution and the expansion rate. The mass
dependence of this suppression can be estimated from the evaporation
timescale, $t_{\rm evap} \propto M^3$, together with the relation between
cosmic time and redshift in the radiation-dominated era, $t \propto
(1+z)^{-2}$. Combining these scalings gives
\begin{equation}
1+z_{\rm eff}(M) \propto M^{-3/2},
\label{eq:zeff_scaling}
\end{equation}
indicating that lighter PBHs evaporate at higher redshifts and therefore
experience stronger redshift dilution. The parametrization $\mathcal{W}(E,M)$
thus encodes both the energy-dependent suppression arising from the redshift
mapping and the mass dependence associated with the evaporation epoch.
While this form is phenomenological, it captures the dominant features of
the full cosmological integration relevant for the present analysis,
and its validity is assessed through a comparison with the
numerical cosmological integral in Sec.~\ref{sec:results}.

To derive constraints on the PBH abundance, we compare the predicted flux
with the observed IceCube spectrum,
\begin{equation}
\Phi_{\rm IC}(E) = \Phi_0 \left(\frac{E}{E_0}\right)^{-\gamma},
\end{equation}
with spectral index $\gamma$ and normalization $\Phi_0$
fixed by the IceCube measurement at reference energy $E_0 = 10^5~{\rm GeV}$.
In this work we consider the IceCube 2020~\cite{IceCube:2020acn} combined analysis 
with $\gamma = 2.37$ as our primary dataset, and assess sensitivity to the
spectral index using the IceCube HESE 2022 result~\cite{IceCube:2020wum} with
$\gamma = 2.87$. We note that the IceCube diffuse flux represents a
measured astrophysical signal rather than a strict upper limit. Our
constraint procedure conservatively requires that any PBH contribution does
not exceed the observed flux, implicitly assuming that the dominant
astrophysical component saturates the measurement. A fully rigorous
treatment would require a joint likelihood analysis incorporating all
neutrino source populations, detector response, and statistical
uncertainties, which is beyond the scope of the present phenomenological
study. Imposing the requirement that the PBH-induced flux does not exceed
the observed IceCube flux at any energy leads to
\begin{equation}
f_{\rm PBH}^{\max}(M,k)
=\min_E
\left[
\frac{\Phi_{\rm IC}(E)}
{\Phi_{\rm PBH}^{(f_{\rm PBH}=1)}(E;M,k)}
\right],
\label{eq:constraints}
\end{equation}
where the minimization is performed over the energy range
$E \in [10^3, 10^6]~{\rm GeV}$ probed by IceCube, and
$\Phi_{\rm PBH}^{(f_{\rm PBH}=1)}$ denotes the predicted flux evaluated at
unit PBH fraction, exploiting the linear scaling $\Phi_{\rm PBH} \propto
f_{\rm PBH}$.

This framework isolates the impact of entropy-driven spectral suppression
and its interplay with cosmological redshift, enabling a transparent
interpretation of how memory-burden effects modify high-energy neutrino
constraints on PBHs.


\section{Results and Implications}
\label{sec:results}

We now present the phenomenological implications of the framework developed
in Sec.~\ref{sec:theory}, focusing on the memory-burden suppression factor,
the diffuse neutrino spectrum, and the resulting constraints on the PBH
dark matter fraction.

\medskip

\begin{figure}[t]
\centering
\includegraphics[width=0.92\textwidth]{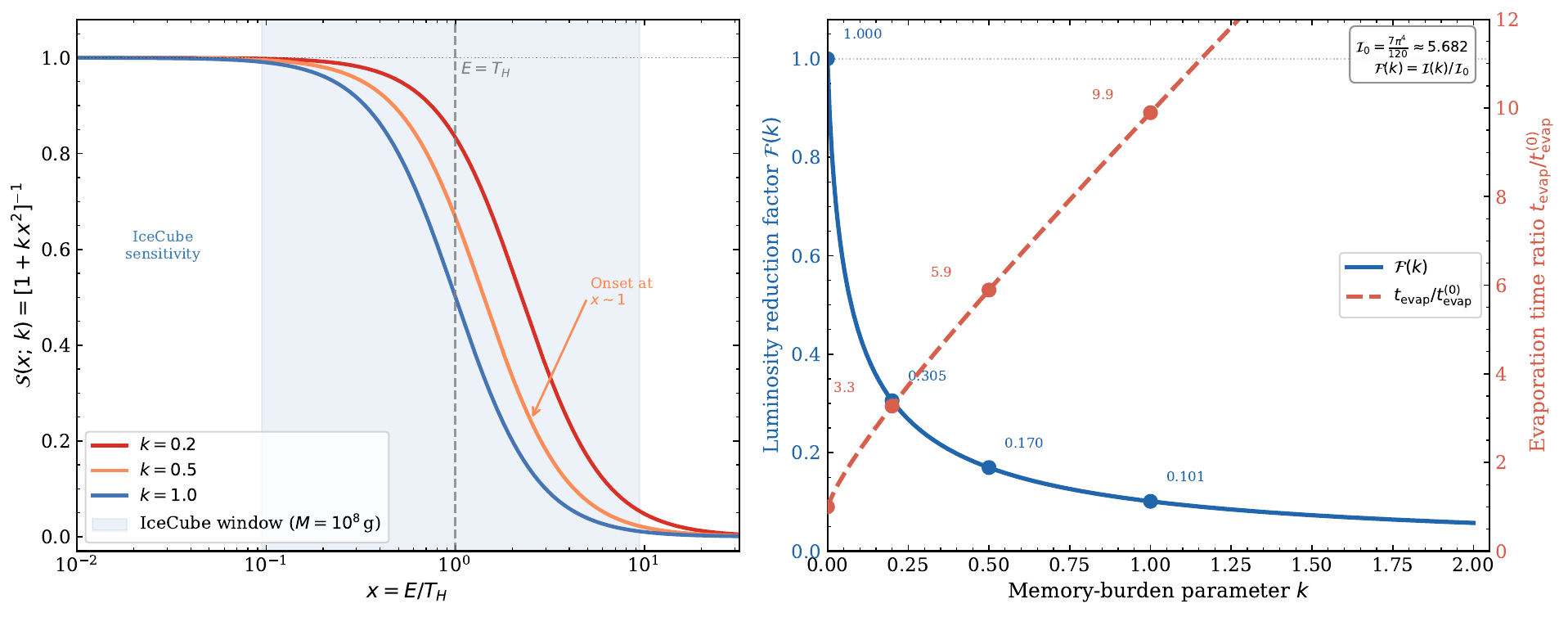}
\caption{Left: memory-burden spectral suppression factor
$\mathcal{S}(x;k) = (1+kx^2)^{-1}$ as a function of $x = E/T_H$ for
$k = 0.2, 0.5, 1.0$. The shaded band marks the IceCube sensitivity window
in $x$-units for $M = 10^8~{\rm g}$. The suppression becomes significant
at $x \gtrsim 1$, placing its onset within the observational window.
Right: luminosity reduction factor $\mathcal{F}(k)$ (blue, left axis) and
evaporation time ratio $t_{\rm evap}/t_{\rm evap}^{(0)} = 1/\mathcal{F}(k)$
(red dashed, right axis) as functions of $k$. Marked points correspond to
the values in Table~\ref{tab:Fk}. For $k = 1$, the total luminosity is
reduced to $\approx 10\%$ of the standard value and the evaporation time
is extended by a factor of $\approx 10$.}
\label{fig:theory}
\end{figure}

\noindent\textbf{Memory-burden suppression and evaporation time:}\\
\\
Before presenting the observable flux, we summarise the
two central theoretical results of Sec.~\ref{sec:theory} in
Fig.~\ref{fig:theory}. The left panel shows the instantaneous spectral
suppression factor $\mathcal{S}(x;k) = (1+kx^2)^{-1}$ as a function of
$x = E/T_H$ for $k = 0.2, 0.5, 1.0$. The suppression is negligible for
$x \ll 1$ (low-energy emission unaffected) and becomes significant for
$x \gtrsim 1$. The shaded band marks the approximate IceCube sensitivity
window in $x$-units for $M = 10^8~{\rm g}$, showing that the onset of
suppression falls squarely within the observational window. This directly
explains why memory-burden effects impact IceCube observables rather than
affecting only asymptotic high-energy emission.

The right panel shows the luminosity reduction factor
$\mathcal{F}(k)$ (blue curve, left axis) and the corresponding evaporation
time enhancement $t_{\rm evap}/t_{\rm evap}^{(0)} = 1/\mathcal{F}(k)$
(red dashed curve, right axis) as functions of $k$. The annotated points
correspond to the numerical values listed in Table~\ref{tab:Fk}. The rapid
decrease of $\mathcal{F}(k)$ with increasing $k$ demonstrates that even
modest memory-burden suppression substantially reduces the total radiated
power and extends the evaporation lifetime: for $k = 1$, the total
luminosity is reduced to approximately $10\%$ of the standard Hawking
value, extending the evaporation time by a factor of $\sim\!10$.

\medskip

\begin{figure}[t]
\centering
\includegraphics[width=0.92\textwidth]{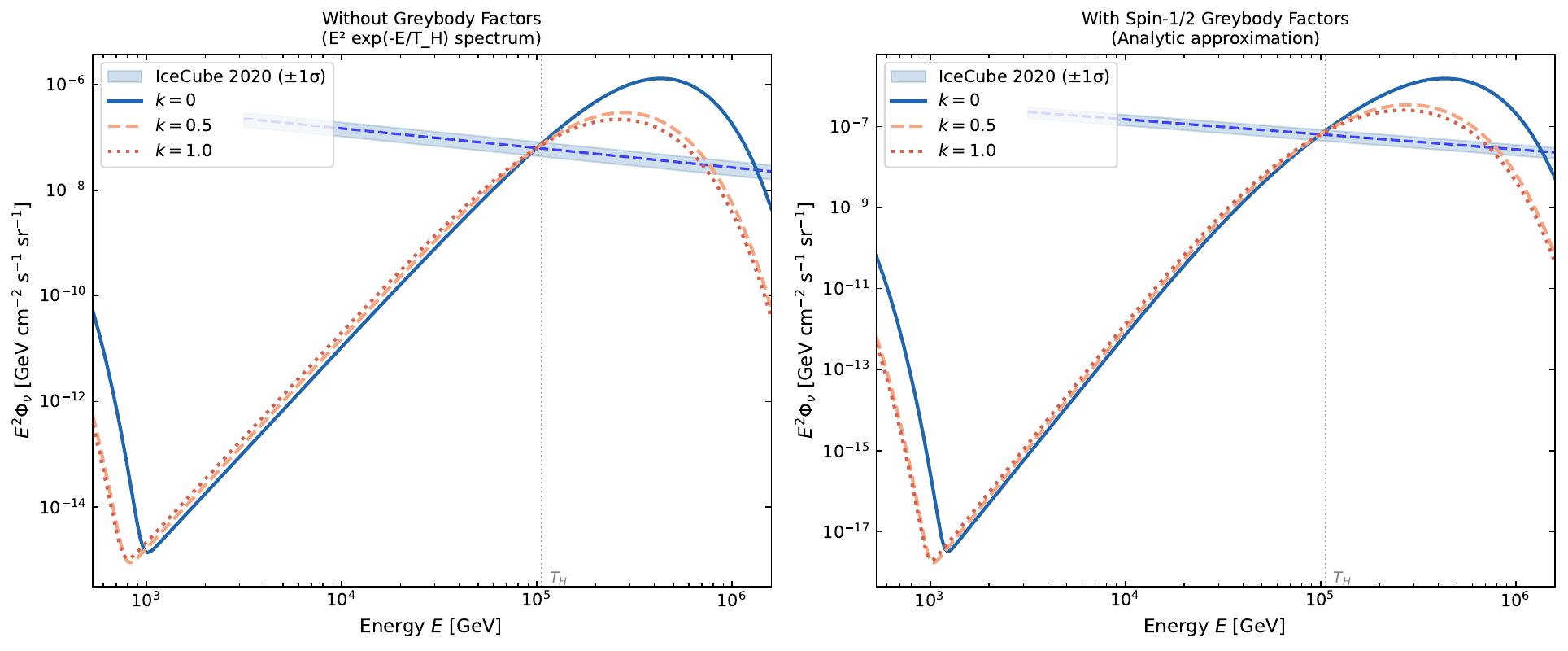}
\caption{Diffuse neutrino flux $E^2\Phi_\nu$ from PBHs as a function of
observed energy for $k = 0, 0.5, 1.0$ and $M = 10^8~{\rm g}$. Left panel:
base Fermi-Dirac spectrum without greybody factors. Right panel: including
the analytic spin-$\frac{1}{2}$ greybody approximation of
Eq.~(\ref{eq:gbf}). The IceCube 2020 measured flux and its
$\pm 1\sigma$ band~\cite{IceCube:2020acn} are shown for comparison. The
vertical dotted line marks $E = T_H \approx 10^5~{\rm GeV}$.
The overall normalisation is treated as effective within
the phenomenological framework (see Sec.~\ref{sec:theory}), and the
vertical separation between the PBH curves and the IceCube band reflects
the order-of-magnitude nature of the flux estimate. The key observable
quantity is the relative suppression between curves of different $k$,
which is independent of the normalisation.}
\label{fig:flux}
\end{figure}

\noindent\textbf{Diffuse neutrino spectrum:}\\
\\
The diffuse neutrino flux from a cosmological PBH population is computed
using Eq.~(\ref{eq:flux_final}). The resulting spectra for representative
values of the memory-burden parameter $k = 0, 0.5, 1.0$ are shown in
Fig.~\ref{fig:flux}, both without (left panel) and with (right panel)
spin-$\frac{1}{2}$ greybody corrections.

The spectrum exhibits a characteristic peak at $E_{\rm peak} \sim
\mathcal{O}(T_H) \propto M^{-1}$, reflecting the thermal nature of Hawking
emission. For $M = 10^8~{\rm g}$, the Hawking temperature is
$T_H \approx 10^5~{\rm GeV}$, and the spectral peak lies at
$E_{\rm peak} \sim {\rm few} \times T_H$, within the IceCube sensitivity
window~\cite{IceCube:2020wum}.

The greybody factor modifies the spectral shape at low
energies $E \ll T_H$, where $\Gamma_\nu \propto x^2 \to 0$ suppresses the
emission relative to the no-GBF case. This is visible as a suppression of
the rising low-energy branch in the right panel compared to the left. At
energies $E \gtrsim T_H$, the greybody factor approaches unity and its
effect is negligible. The overall normalisation differs between the two
panels because the GBF reduces the total integrated flux; however, as we
discuss below, this order-of-magnitude difference does not affect the
relative memory-burden suppression, which is the physically meaningful
quantity for constraining $f_{\rm PBH}$.

As $k$ increases, the high-energy tail is suppressed according to
Eq.~(\ref{eq:S_final}), reducing the overlap with the IceCube sensitivity
band and consequently weakening the observable signal. The suppression is
visible in both panels as a progressive reduction of the flux above
$E \sim T_H$, with the low-energy portion of the spectrum remaining
unaffected in both cases.

The vertical separation between the PBH flux curves and the IceCube band in Fig.~\ref{fig:flux} 
reflects the order-of-magnitude nature of the effective normalization adopted in Eq.~(\ref{eq:flux_final}). 
The overall amplitude of the predicted flux depends on the time-integrated emission spectrum $dN/dE$, 
the effective redshift factor $\mathcal{W}(E,M)$, and the assumed PBH fraction $f_\mathrm{PBH}$, 
none of which are fixed independently within the present framework. A rough estimate of the expected 
flux magnitude can be obtained by evaluating the cosmological integral in Eq.~(\ref{eq:full_flux}) 
at $f_\mathrm{PBH} = 1$ and $M = 10^8\,\mathrm{g}$: using $\rho_\mathrm{DM,0} \simeq 
2.4 \times 10^{-30}\,\mathrm{g\,cm^{-3}}$, $t_\mathrm{evap} \sim 10^{-15}\,\mathrm{s}$, and 
the cosmological volume factor $c/H_0 \sim 10^{28}\,\mathrm{cm}$, one obtains 
$E^2\Phi \sim 10^{-8}$--$10^{-6}\,\mathrm{GeV\,cm^{-2}\,s^{-1}\,sr^{-1}}$ before redshift 
suppression. The additional suppression from $\mathcal{W}(E,M)$, which encodes both the 
$(1+z_\mathrm{eff})^{-1}$ dilution and the energy-dependent factor $(E/E_*)^\alpha$, accounts 
for the remaining gap. The key physical content of Fig.~\ref{fig:flux} is therefore 
the relative suppression between curves of different $k$, which is independent of this 
normalization uncertainty and is the quantity that directly enters the constraint analysis.

\medskip

\begin{figure}[t]
\centering
\includegraphics[width=0.92\textwidth]{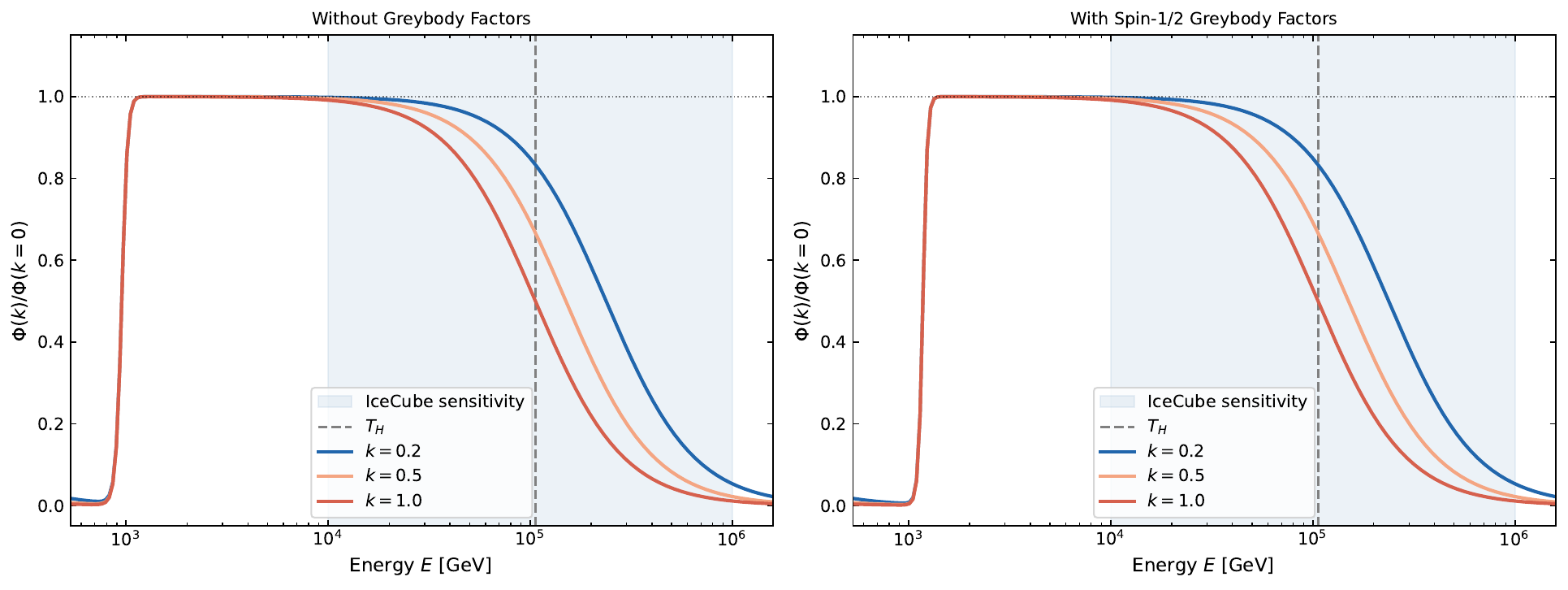}
\caption{Spectral ratio $\mathcal{R}(E;k) = \Phi(k)/\Phi(k=0)$ as a
function of observed energy, without (left) and with (right) greybody
corrections, for $M = 10^8~{\rm g}$.
The vertical dashed line marks $E = T_H$ and the shaded region denotes
the approximate IceCube sensitivity window ($10^4$--$10^6~{\rm GeV}$).
The ratio $\mathcal{R}(E;k)$ is independent of the
overall flux normalisation and of the greybody factor, which cancels
between numerator and denominator. The two panels are therefore nearly
identical by construction; any residual difference arises only at
$E \ll T_H$ where the greybody factor has a non-trivial energy dependence.
The suppression onset at $E \sim T_H$ falls within the IceCube window,
demonstrating that memory-burden effects directly reduce the observable
signal rather than acting only at asymptotically high energies.}
\label{fig:ratio}
\end{figure}

\noindent\textbf{Spectral deformation from memory burden:}\\
\\
To isolate the spectral modification due to memory burden independently of
the overall normalisation, we compute the ratio
\begin{equation}
\mathcal{R}(E;k) \equiv \frac{\Phi(E;k)}{\Phi(E;k=0)},
\label{eq:ratio}
\end{equation}
shown in Fig.~\ref{fig:ratio} for $k = 0.2, 0.5, 1.0$.

The ratio $\mathcal{R}(E;k)$ is close to unity for $E \ll T_H$,
confirming that the low-energy emission is unaffected by memory burden.
The suppression becomes significant for $E \gtrsim T_H$, as dictated by
the functional form of $\mathcal{S}(E,M;k)$ in Eq.~(\ref{eq:S_final}),
and increases monotonically with both $E$ and $k$.

An important observation concerns the two panels of
Fig.~\ref{fig:ratio}: they are nearly identical. This is not a numerical
coincidence but a mathematical consequence of the fact that the greybody
factor $\Gamma_\nu(E,M)$ multiplies the spectrum equally for all values
of $k$, and therefore cancels exactly in the ratio $\mathcal{R}(E;k)$.
The greybody factor modifies the absolute amplitude of $\Phi$ but not the
relative memory-burden suppression. This confirms that the spectral
deformation quantified by $\mathcal{R}(E;k)$ is a robust probe of memory
burden, insensitive to uncertainties in the emission model.

Since the suppression onset occurs within the IceCube sensitivity window,
memory-burden effects directly reduce the detectable flux rather than
acting only at energies far above the observational range. For $k = 1.0$,
the ratio $\mathcal{R} \lesssim 0.5$ throughout most of the IceCube band,
implying a reduction of the observable signal by at least a factor of two
relative to the standard Hawking case.

\medskip

\begin{figure}[t]
\centering
\includegraphics[width=0.92\textwidth]{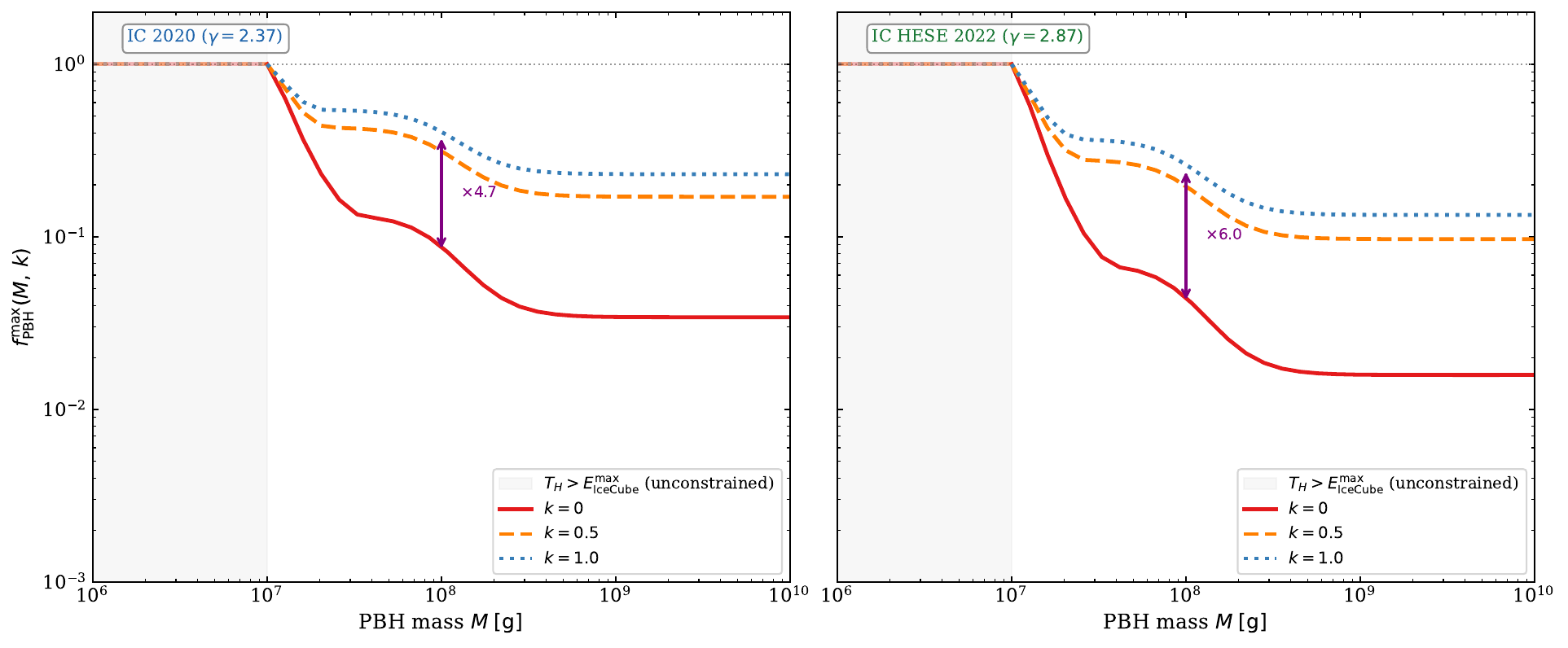}
\caption{Upper bound on the PBH dark matter fraction,
$f_{\rm PBH}^{\max}(M,k)$, as a function of PBH mass $M$ for
$k = 0$ (solid red), $k = 0.5$ (dashed orange), and $k = 1.0$ (dotted
blue). Left panel: IceCube 2020 dataset ($\gamma = 2.37$)~\cite{IceCube:2020acn}.
Right panel: IceCube HESE 2022 dataset ($\gamma = 2.87$)~\cite{IceCube:2020wum}.
The gray shaded region ($M \lesssim 10^7~{\rm g}$) indicates the range
where $T_H$ exceeds the maximum IceCube energy, so that no meaningful
constraint is derived. The double-headed arrows mark the
weakening of the constraint at $M \approx 10^8~{\rm g}$ when going from
$k=0$ to $k=1$: a factor of $\approx 4.7$ (IC 2020) and $\approx 6.0$
(HESE 2022). These figures are consistent with the luminosity reduction
factor $1/\mathcal{F}(k=1) \approx 9.9$ from Table~\ref{tab:Fk}, with the
partial offset arising from the competing evaporation delay effect discussed
in Sec.~\ref{sec:theory}.}
\label{fig:constraint}
\end{figure}

\noindent\textbf{Constraints on PBH abundance:}\\
\\
We now derive constraints on the PBH dark matter fraction using
Eq.~(\ref{eq:constraints}). The resulting upper bounds $f_{\rm PBH}^{\max}$
are shown in Fig.~\ref{fig:constraint} as a function of PBH mass, for
$k = 0, 0.5, 1.0$ and for both the IceCube 2020~\cite{IceCube:2020acn} and
HESE 2022~\cite{IceCube:2020wum} datasets.

The constraint curves display a qualitatively distinct
structure compared to the standard expectation and we describe the three
regimes explicitly.

\textit{Unconstrained regime} ($M \lesssim 10^7~{\rm g}$,
gray shaded region): For these masses, the Hawking temperature satisfies
$T_H \gtrsim 10^6~{\rm GeV}$, which exceeds the upper boundary of the
IceCube sensitivity window. The PBH spectrum peaks above the detector
range, and even accounting for the redshift suppression parametrized by
$\mathcal{W}(E,M)$, no meaningful constraint on $f_{\rm PBH}$ can be
derived within the present framework.

\textit{Transition regime} ($10^7 \lesssim M \lesssim
{\rm few} \times 10^7~{\rm g}$): As the mass increases, $T_H$ enters the
upper portion of the IceCube window. The constraints strengthen rapidly,
dropping from $f_{\rm PBH}^{\max} = 1$ to values of order $10^{-1}$
within less than a decade in mass.

\textit{Constrained plateau} ($M \gtrsim
{\rm few} \times 10^7~{\rm g}$): For larger masses, the constraints flatten
to an approximate plateau. This reflects the fact that as $T_H$ decreases
through the IceCube window and eventually falls below $10^3~{\rm GeV}$, the
normalization procedure compensates by increasing the required $f_{\rm PBH}$.
The competing effects of decreasing $T_H$ and decreasing $z_{\rm eff}$
(which reduces the $\mathcal{W}$ suppression) partially cancel, producing a
near-flat constraint in the range $M \sim 10^8$--$10^{10}~{\rm g}$, with
values $f_{\rm PBH}^{\max}(k=0) \approx 3$--$8 \times 10^{-2}$ for IC 2020.

It is instructive to compare the $k = 0$ baseline with existing neutrino-based constraints 
on the PBH dark matter fraction derived under the standard Hawking spectrum. 
Lunardini and Perez-Gonzalez~\cite{Lunardini:2019zob} and Bernal et al.~\cite{Bernal:2022swt} 
derived IceCube constraints in the mass range $M \sim 10^7$--$10^{10}\,\mathrm{g}$, obtaining 
$f_\mathrm{PBH} \lesssim \mathcal{O}(10^{-2}\text{--}10^{-1})$ at comparable masses, 
broadly consistent with our $k = 0$ curves. Dasgupta, Laha, and Ray~\cite{Dasgupta:2019cae} 
similarly found constraints of order $f_\mathrm{PBH} \lesssim 10^{-2}$ at $M \sim 10^8\,\mathrm{g}$ 
using a full cosmological integration. The quantitative agreement at the order-of-magnitude 
level provides a consistency check on the effective framework adopted here, and confirms that 
the absolute normalization uncertainty does not compromise the relative constraint-weakening 
factors derived for $k > 0$, which are the central results of this work. Residual differences 
at the factor-of-few level are expected given the simplified redshift treatment and the 
absence of full numerical greybody factors in the present analysis.

The inclusion of memory-burden effects shifts the constraint curves upward
across the entire mass range, indicating weaker bounds on $f_{\rm PBH}$.
At $M \approx 10^8~{\rm g}$, the constraint weakens by a
factor of $\approx 4.7$ (IC 2020) and $\approx 6.0$ (HESE 2022) when
increasing $k$ from $0$ to $1$. These factors are directly annotated in
Fig.~\ref{fig:constraint}. Notably, the weakening factor is slightly smaller
than the evaporation time ratio $1/\mathcal{F}(k=1) \approx 9.9$ listed in
Table~\ref{tab:Fk}. This is because the constraint weakening is driven by
the spectral suppression $\mathcal{S}(E;k)$ acting on the observable flux,
while the evaporation time enhancement $1/\mathcal{F}(k)$ reflects the
suppression of the \emph{total} integrated luminosity. The two are related
but not identical: the observable flux at a fixed reference energy receives
contributions from a range of $x = E/T_H$ values, and the effective
suppression is therefore an average of $\mathcal{S}(x;k)$ over the IceCube
window rather than the full integral $\mathcal{F}(k)$.

Comparing the two panels, the HESE 2022 dataset with
steeper spectral index $\gamma = 2.87$ yields somewhat stronger constraints
than IC 2020, particularly at lower masses where the IceCube flux falls more
steeply and the PBH contribution is relatively more significant. The
qualitative structure of the constraints and the direction and magnitude of
the memory-burden weakening are consistent between the two datasets,
demonstrating the robustness of the conclusions with respect to the choice
of IceCube measurement.

A numerical summary of representative constraint values is given in
Table~\ref{tab:constraints}.

\textcolor{blue}{
\begin{table}[h]
\centering
\caption{Upper bounds $f_{\rm PBH}^{\max}(M,k)$ at
selected masses for the IC 2020 dataset ($\gamma = 2.37$). The final column
gives the weakening factor relative to $k = 0$.}
\label{tab:constraints}
\begin{tabular}{ccccc}
\hline\hline
$M\,[\mathrm{g}]$ & $f_{\rm PBH}^{\max}(k{=}0)$ &
$f_{\rm PBH}^{\max}(k{=}0.5)$ & $f_{\rm PBH}^{\max}(k{=}1.0)$ &
$f_{\rm max}(k{=}1)/f_{\rm max}(k{=}0)$ \\
\hline
$3\times 10^7$ & $1.3\times 10^{-1}$ & $4.2\times 10^{-1}$ &
$5.4\times 10^{-1}$ & $4.0$ \\
$10^8$         & $8.2\times 10^{-2}$ & $3.0\times 10^{-1}$ &
$3.9\times 10^{-1}$ & $4.7$ \\
$3\times 10^8$ & $3.9\times 10^{-2}$ & $1.9\times 10^{-1}$ &
$2.5\times 10^{-1}$ & $6.3$ \\
$10^9$         & $3.4\times 10^{-2}$ & $1.7\times 10^{-1}$ &
$2.3\times 10^{-1}$ & $6.7$ \\
\hline\hline
\end{tabular}
\end{table}
}


\section{Conclusion}
\label{sec:conc}

In this work, we have investigated the impact of quantum gravitational
memory-burden effects on neutrino signals from evaporating primordial black
holes (PBHs) and the corresponding constraints from IceCube observations.
We adopted a phenomenological parametrisation of the entropy-induced spectral
suppression, characterised by a dimensionless parameter $k$, and derived the
implications for the diffuse neutrino flux and the PBH dark matter fraction.

Our central theoretical result is the derivation of the
memory-burden modified evaporation time, Eq.~(\ref{eq:tevap}):
\begin{equation*}
t_{\rm evap}(M_0, k) = \frac{M_0^3}{3\,\alpha\,\mathcal{F}(k)}
= \frac{t_{\rm evap}^{(0)}}{\mathcal{F}(k)},
\end{equation*}
where the luminosity reduction factor $\mathcal{F}(k)$ is given by the
dimensionless integral in Eq.~(\ref{eq:Fk}), evaluated using the
Fermi-Dirac emission spectrum consistent with the fermionic nature of
neutrinos. The key property that makes this result analytically exact within
the present framework is that $\mathcal{F}(k)$ depends only on $k$ and not
on $M$ or $T_H$, so the modified mass-loss equation retains the same
functional form as the standard Hawking case and can be integrated
analytically. For $k = 0.5$ and $k = 1.0$, the evaporation time is extended
by factors of $\approx 5.9$ and $\approx 9.9$ respectively.

We have shown that memory-burden effects primarily suppress the high-energy
tail of the instantaneous emission spectrum, while leaving the low-energy
component largely unaffected. The onset of this suppression occurs at
$E \sim T_H$, which falls within the IceCube sensitivity window for PBH
masses $M \sim 10^7$--$10^8~{\rm g}$. As a result, memory burden directly
reduces the overlap between the PBH neutrino spectrum and the detector
sensitivity band, leading to a systematic weakening of the derived
constraints on the PBH dark matter fraction.

The spectral ratio $\mathcal{R}(E;k) = \Phi(k)/\Phi(k=0)$,
shown in Fig.~\ref{fig:ratio}, provides a normalisation-independent and
model-robust measure of this suppression. We have explicitly demonstrated
that $\mathcal{R}(E;k)$ is insensitive to the inclusion of greybody factors,
since these cancel in the ratio, and is therefore a clean diagnostic of the
memory-burden effect alone.

Using the phenomenological effective redshift framework of
Eq.~(\ref{eq:flux_final}), we derived upper bounds $f_{\rm PBH}^{\max}(M,k)$
on the PBH dark matter fraction by comparing the predicted flux with the
observed IceCube diffuse neutrino spectrum. The constraint
analysis reveals three qualitatively distinct regimes: an unconstrained
regime for $M \lesssim 10^7~{\rm g}$ (where $T_H$ exceeds the IceCube
energy range), a rapid transition around $M \sim {\rm few}\times 10^7~{\rm
g}$, and an approximate plateau at larger masses. The plateau structure
arises from the partial cancellation between decreasing Hawking temperature
and decreasing effective redshift suppression as the mass increases.

The inclusion of memory-burden effects weakens the constraints across the
entire constrained mass range. At $M \approx 10^8~{\rm g}$,
the constraint on $f_{\rm PBH}$ weakens by a factor of $\approx 4.7$ (IC
2020) to $\approx 6.0$ (HESE 2022) when increasing $k$ from $0$ to $1$.
This weakening factor is somewhat smaller than the evaporation time
enhancement $1/\mathcal{F}(k=1) \approx 9.9$, because the constraint is
determined by the spectral suppression $\mathcal{S}(E;k)$ averaged over the
IceCube window rather than the total integrated luminosity. The results are
qualitatively consistent between the IC 2020 and HESE 2022 datasets, with
the steeper spectral index of the latter yielding modestly stronger
constraints.

These results should be interpreted within the approximations adopted in
this work. The use of a simplified Fermi-Dirac emission spectrum, an
effective redshift parametrisation $\mathcal{W}(E,M)$, and a flux-matching
normalisation procedure at a reference energy introduces order-of-magnitude
uncertainties in the absolute constraint values. A more rigorous treatment
incorporating full numerical greybody factors via a code such as
\textsc{BlackHawk}~\cite{Arbey:2019mbc,Arbey:2021mbl}, a detailed
cosmological evolution of the PBH population, and a likelihood-based
comparison with IceCube data would refine the quantitative bounds.
However, the qualitative conclusions --- that memory-burden
effects suppress the high-energy neutrino flux and systematically weaken the
derived constraints --- are expected to persist under such improvements,
since they follow directly from the spectral modification $\mathcal{S}(E;k)$
and are independent of the normalization procedure.

Future work may extend this framework in several directions: implementing
full numerical cosmological calculations with memory-burden modified
evaporation rates, exploring the interplay between the spectral suppression
and the modified evaporation redshift in the mass range where the two effects
are comparable ($M \sim 10^{13}$--$10^{14}~{\rm g}$), and investigating
complementary multi-messenger signatures in gamma rays and gravitational
waves. The modified evaporation lifetime derived here also has direct
implications for the formation of long-lived PBH remnants, which constitute
a distinct dark matter candidate~\cite{Alexandre:2024nuo,Dvali:2024hsb} and
merit separate investigation.


\bibliography{ref}
\end{document}